\documentclass[aps,prl,twocolumn,groupedaddress,showpacs]{revtex4}
\begin{document}

\newcommand{\ket}[1]{|#1 \rangle}
\newcommand\reals{\rm R}
\newcommand\integers{\rm Z}

\preprint{APS/123-QED}

\title{Eigenvector Approximation Leading to Exponential Speedup
of Quantum Eigenvalue Calculation}

\author{Peter Jaksch}
\email{petja@cs.columbia.edu}
\author{Anargyros Papageorgiou}
\email{ap@cs.columbia.edu}
\affiliation{Department of Computer Science, Columbia University.}

\date{\today}

\begin{abstract}
We present an efficient method for preparing the initial state required 
by the eigenvalue approximation quantum algorithm of 
Abrams and Lloyd \cite{AL99}. 
Our method can be applied when solving continuous Hermitian eigenproblems, 
e.g., the Schr\"odinger equation, on a discrete grid. We start 
with a classically obtained eigenvector for a problem discretized on a coarse
grid, and we efficiently construct, quantum mechanically, an approximation of 
the same eigenvector on a fine grid. 
We use this approximation as the initial state for the eigenvalue estimation algorithm, and show the relationship between its success
probability and the size of the coarse grid.
\end{abstract}


\maketitle

Intuitively, quantum mechanical problems offer great potential for quantum
computers to achieve large speedups over classical machines. An important 
problem of this kind is approximation of an eigenvalue of a quantum mechanical
evolution operator.
In a recent paper \cite{AL99}, Abrams and Lloyd present a quantum 
algorithm for doing this. Their algorithm is exponentially faster than
the best classical algorithm, but requires a good approximation of an 
eigenvector as input. In this paper we show how to obtain an
approximation efficiently which is guaranteed to be good.

The key component of the algorithm in \cite{AL99} is quantum phase estimation, 
which is a method for approximating an eigenvalue of a unitary 
matrix \cite{nielsen}. We give a brief outline of this algorithm below.

Let $Q$ denote a $2^m \times 2^m$ unitary matrix. We want to approximate
a specific eigenvalue of $Q$. Phase estimation does this using the 
corresponding eigenvector as input. 
The algorithm in \cite{AL99} deals with the case when this eigenvector is not
known exactly. In particular, consider a quantum computer consisting of 
three registers with a total of $b+m+w$ qubits. The first $b$ qubits are all 
initially in the state $\ket{0}$. The second register with $m$ qubits is 
initialized to some state $\ket{\psi}$, which must approximate the 
eigenvector in question sufficiently well, as we will see. 
The last $w$ qubits are work qubits for temporary storage. 

Since $Q$ is unitary and therefore normal, the state $\ket{\psi}$ can be 
expanded with respect to eigenvectors of $Q$. Omitting the work qubits, 
the initial state of the algorithm is
\begin{equation}
\ket{0}\ket{\psi} = \ket{0} \sum_{u}d_u\ket{u} ,
\end{equation}
where $\ket{u}$ are the eigenvectors of $Q$. Placing the first register in an equal superposition, using $b$ Hadamard gates, transforms this state into
\begin{equation}
\frac{1}{\sqrt{2^b}}\sum_{j=0}^{2^b-1}\ket{j} \sum_{u}d_u\ket{u} .
\end{equation}
Next, powers of $Q$ are applied to create the state
\begin{equation}
\frac{1}{\sqrt{2^b}}\sum_{j=0}^{2^b-1}\ket{j} Q^j  \sum_{u}d_u\ket{u}. \label{Uj}
\end{equation}
Since $Q$ is unitary, its eigenvalues can be written as 
$e^{2\pi i\varphi_u}$, where $\varphi_u \in \reals$.
We can assume that $\varphi_u \in [0,1)$ and consider the approximation 
of one of these phases instead of the approximation of one of the eigenvalues.
Equation (\ref{Uj}) is equal to 
\begin{equation}
\frac{1}{\sqrt{2^b}} \sum_{u} \sum_{j=0}^{2^b-1}  d_u e^{2\pi ij\varphi_u}\ket{j} \ket{u}.
\end{equation}
It is easily seen that the inverse Fourier transform performed on the first register creates the state
\begin{equation}
\sum_{u} d_u \left( \sum_{j=0}^{2^b-1} g(\varphi_u,j) 
\ket{j} \right) \ket{u}, \label{invfourier}
\end{equation}
where 
\begin{equation}
g(\varphi_u,j) = \left\{ \begin{array}{ll} \frac{\sin (\pi(2^b\varphi_u -j))
e^{\pi i(\varphi_u - j2^{-b})(2^b-1)}} {2^b\sin(\pi (\varphi_u-j2^{-b}))}, 
& 2^b\varphi_u \neq j \\
1, & 2^b\varphi_u = j.
\end{array} \right.
\end{equation}
A measurement of the first register produces outcome $j$ with probability
\begin{equation} 
p_j = \sum_u |d_u|^2|g(\varphi_u,j)|^2,
\end{equation}
and the second register will collapse to the state
\begin{equation}
\sum_{u} \frac{d_ug(\varphi_u,j)}{\sqrt{p_j}} \ket{u}. \label{eigensuper}
\end{equation}
We remark that for the special case when the eigenvalues $\varphi_u$ can be 
represented exactly with $b$-bits (i.e., $2^b\varphi_u$ is an integer), 
equation (\ref{invfourier}) simplifies to
\begin{equation}
\sum_u d_u \ket{\varphi_u} \ket{u}.
\end{equation} 
Thus, when the eigenvalues are of this form an distinct a measurement of the 
first register will cause the second register to collapse exactly onto the 
corresponding eigenvector.

Recall that we are interested in approximating the phase that corresponds
to an eigenvector $\ket{u'}$, that the state $\ket{\psi}$ is an 
approximation of this eigenvector and that the eigenvalue is of the form
$e^{2\pi i \varphi_{u'}}$. For instance, one is often interested in the 
eigenvalue corresponding to the ground state. We define 
$\Delta(\varphi_0,\varphi_1) = \min_{x \in \integers}\{|x + \varphi_1 - \varphi_0|\}$,
$\varphi_0, \varphi_1 \in \reals$ (i.e., the fractional part of the distance
between $\varphi_0$ and $\varphi_1$). 
Then a measurement of the first register produces an outcome from the set 
$\mathcal{G} = \{ j: \Delta(j/2^b,\varphi_{u'}) \leq k/2^b,
\quad k>1 \}$ with probability
\begin{eqnarray}
\Pr (\mathcal{G}) &=& 
	\sum_{j \in \mathcal{G}} \sum_u |d_ug(\varphi_u,j)|^2 \nonumber \\
&\ge& \sum_{j \in \mathcal{G}} |d_{u'}g(\varphi_{u'},j)|^2 \nonumber \\
&\ge& |d_{u'}|^2 - \frac{|d_{u'}|^2}{2(k-1)},
\label{Pj}
\end{eqnarray}
and when $k=1$ the probability that $\Delta(j/2^b,\varphi_{u'}) \leq 2^{-b}$ is
bounded from below by $\frac{8}{\pi^2}|d_{u'}|^2$; 
where the proofs of the probability bounds can be found in \cite{Brassard,nielsen}. Observe that $\ket{\psi}$ must be chosen in a way that this 
probability is greater than $\frac{1}{2}$, which implies that 
$|d_{u'}|$ has to be sufficiently large. If we want to obtain an approximation of $\varphi_{u'}$ with accuracy $2^{-n}$ and
probability at least $|d_{u'}|^2(1-\epsilon)$, equation (\ref{Pj}) shows
that this can achieved by choosing the number of qubits $b$ in the first 
register to be
\begin{equation}
b = n + \left\lceil \log \left(1+\frac{1}{2\epsilon} \right) \right\rceil.
\end{equation}

The algorithm in \cite{AL99} is based on the fact that 
quantum phase estimation can be used as an efficient subroutine to find 
eigenvalues. Consider a Hermitian operator of the form 
$H = \sum H_j$, where each $H_j$ acts only on a small number 
of qubits. Since $H$ is Hermitian the operator $G(t)=e^{-iHt}$ is unitary 
and has the same eigenvectors as $H$. Using the technique in~\cite{lloyd96}, 
$G(t)$ can be approximated efficiently 
on a quantum computer and the approximation is used as the unitary operator in 
the quantum phase estimation algorithm. (For more details regarding the efficient implementation of $G$ see \cite{AL99} and the references therein.)

The Hermitian eigenproblem described above is solved  on a discrete grid.  
We are interested in the case when the grid is extremely fine. Clearly, a 
fine grid requires a large vector for the representation of the initial 
state of the algorithm. In general, it may not be possible to 
efficiently prepare an arbitrary quantum state in a space with a large number 
of qubits. However, in our case we will show a method for the efficient 
preparation of an initial state.

Suppose we have an eigenvector for a coarse grid discretization of the problem.
We can assume that we obtained it classically because the size of the problem
is small.
Using this eigenvector, we efficiently construct 
an approximation to the corresponding eigenvector for a fine grid 
discretization of the problem. 
We use this approximation as the initial state of the eigenvalue
 approximation algorithm.
We describe our method for a one-dimensional continuous problem on 
the interval $[0,1]$. 

Let $H$ be an positive Hermitian operator, defined 
on a Hilbert space of smooth functions on $[0,1]$. Let $v_k(\cdot)$,
$k=1,2,\dots$, 
denote the eigenfunctions of $H$, ordered according to the magnitude of the
corresponding eigenvalues; and without loss of generality we assume that 
\begin{equation}
\int_0^1 |v_k(x)|^2 dx = 1. \label{normalization}
\end{equation}
Suppose that $H_N$ is a discretization of $H$ with grid size $h_N=1/(1+N)$. 
Let $\ket{U_k^{(N)}}$, $k=0,1,\ldots,N-1$, denote the normalized 
eigenvectors of $H_N$, ordered according to the magnitude of the 
corresponding eigenvalues. The expansion of the $k$-th eigenvector in the 
computational basis can be written as 
\begin{equation}
\ket{U_k^{(N)}} = \sum_{j=0}^{N-1} u_{k,j}^{(N)} \ket{j}. \label{compexpansion}
\end{equation}
Let $\ket{V_k^{(N)}} = \sum_{j=0}^{N-1} v_k\left((j+1)h_N\right) \ket{j}$ 
be the sampled version of $v_k(\cdot)$ at the discretization points.
Consider problems such that the eigenvector of interest satisfies
$\|v_k^\prime\|_\infty = \sup_{0 \leq x \leq 1} |v_k^\prime(x)| = O(1)$ and
\begin{equation}
\left\| \ket{U_k^{(N)}} - \frac {\ket{V_k^{(N)}}}{\| \ket{V_k^{(N)}} \|}\right\| 
= O(h_N^q), \label{pointwise}
\end{equation} 
where $q>0$ is the order of convergence and $\|\ket{X}\|^2 = \sum_{j=0}^{N-1} 
|x_j|^2$, for $\ket{X}=\sum_{j=0}^{j=N-1} x_j \ket{j}$. 
For example, these conditions are satisfied when we are dealing with second 
order elliptic operators; see \cite{Babuska} for the solution of eigenvalue 
problems.

Now, assume that the eigenvector $\ket{U_k^{(N_0)}}$ 
of $H_{N_0}$ has been obtained classically \footnote{In fact, we can weaken this 
condition by considering the numerical error in solving the coarsely 
discretized  problem. It suffices to assume that we have an approximation
$\ket{\hat{U}_k^{(N_0)}}$ of the eigenvector $\ket{U_k^{(N_0)}}$ such that
$\| \ket{\hat{U}_k^{(N_0)}} - \ket{U_k^{(N_0)}} \| = O(h_{N_0}^q)$.}. 
This vector is placed in a $\log N_0$ qubit register. 
For $N=2^sN_0$, we construct an approximation $\ket{\tilde{U}_k^{(N)}}$ 
of $\ket{U_k^{(N)}}$ by appending $s$ qubits, in the state $\ket{0}$, 
to $\ket{U_k^{(N_0)}}$ and then performing a Hadamard transformation on 
each one of these $s$ qubits, i.e.
\begin{eqnarray}
 \ket{\tilde{U}_k^{(N)}} &=&  \ket{U_k^{(N_0)}} \left( \frac{\ket{0}+\ket{1}}{\sqrt{2}} \right)^{\otimes s} \nonumber \\ 
&=&  \frac{1}{\sqrt{2^s}} \sum_{j=0}^{N-1} u_{k,f(j)}^{(N_0)} \ket{j},
\end{eqnarray} 
where $f(j) = \lfloor j/2^k \rfloor$. 
The effect of $f$ is to replicate the coordinates of $\ket{U_k^{(N_0)}}$ $2^s$
times. We use $\ket{\tilde{U}_k^{(N)}}$ as input to the eigenvalue and 
eigenvector approximation algorithm.
When the result of the algorithm is measured $\ket{\tilde{U}_k^{(N)}}$ will 
collapse onto a superposition of eigenvectors according to equation 
(\ref{eigensuper}). 
We show that the magnitude of the coefficient of $\ket{U_k^{(N)}}$ in this
superposition can be made arbitrarily close to one by appropriately 
choosing $N_0$. 

Consider two different expansions of $\ket{\tilde{U}_k^{(N)}}$:
\begin{eqnarray}
\ket{\tilde{U}_k^{(N)}} &=& \sum_{j=0}^{N-1} \tilde{u}_{k,j}^{(N)} \ket{j} \label{approxexpansion}\\
\ket{\tilde{U}_k^{(N)}} &=& \sum_{l=0}^{N-1} d_{k,l}^{(N)} \ket{U_l^{(N)}}. \label{eigenexpansion}
\end{eqnarray}
The first expansion is in the computational basis and the second is with 
respect to the eigenvectors of $H_N$. 
We call $|d_{k,k}^{(N)}|^2$ the probability of success. Equation 
(\ref{eigenexpansion}) can be rewritten as 
\begin{equation} 
\ket{\tilde{U}_k^{(N)}} - \ket{U_k^{(N)}} = (d_{k,k}^{(N)} - 1)\ket{U_k^{(N)}}
+ \sum_{l \neq k} d_{k,l}^{(N)} \ket{U_l^{(N)}}.
\end{equation} 
Taking norms on both sides and using (\ref{compexpansion}) and (\ref{approxexpansion}) gives the inequality 
\begin{eqnarray}
\left| \left| \ket{U_k^{(N)}} - \ket{\tilde{U}_k^{(N)}} \right| \right|^2 &=&  \sum_{j=0}^{N-1} |u_{k,j}^{(N)} - \tilde{u}_{k,j}^{(N)}|^2 \nonumber \\
&=& |d_{k,k}^{(N)}-1|^2 + \sum_{l \neq k} |d_{k,l}^{(N)}|^2 \nonumber \\
&\geq& \sum_{l \neq k} |d_{k,l}^{(N)}|^2 \nonumber \\
&=& 1-|d_{k,k}^{(N)}|^2. \label{errorbound}
\end{eqnarray}

We will now bound (\ref{errorbound}) from above, and thus the probability of 
failure. The definition of $\ket{\tilde{U}_k^{(N)}}$ implies
\begin{eqnarray}
&& \left\| \ket{U_k^{(N)}} - \ket{\tilde{U}_k^{(N)}} \right\|^2
= \sum_{j=0}^{N-1} \Biggl| \frac{v_k((j+1)h_{N})}{\| \ket{V_k^{(N)}} \|} \Biggr. 
\nonumber \\
&& \quad - \frac{v_k((f(j)+1)h_{N_0})}{\sqrt{2^s} \| \ket{V_k^{(N_0)}} \|}  + 
\Delta_{k,j}^{(N)} - \frac{\Delta_{k,f(j)}^{(N_0)}}{\sqrt{2^s}} \Biggl. \Biggr|^2, \label{bound}
\end{eqnarray}
where $\sum_{j=0}^{N-1} |\Delta_{k,j}^{(N)}|^2 = O(h_N^{2q})$ and 
$\sum_{j=0}^{N-1} |\Delta_{k,f(j)}^{(N_0)}|^2 = 2^s O(h_{N_0}^{2q})$
by (\ref{pointwise}). Applying the triangle inequality, we get
\begin{eqnarray}
&& \left\| \ket{U_k^{(N)}} - \ket{\tilde{U}_k^{(N)}} \right\| \leq \left( \sum_{j=0}^{N-1} \left| \frac{v_k((j+1)h_{N})}{\| \ket{V_k^{(N)}} \|}  \right. \right. 
\nonumber \\
&& \quad - \left. \left.  \frac{v_k((f(j)+1)h_{N_0})}{\sqrt{2^s} \| \ket{V_k^{(N_0)}} \|}  \right|^2 \right)^{1/2} + O(h_{N_0}^q).   \label{sumestimate}
\end{eqnarray}
The definition of $\ket{V_k^{(N)}}$ and the fact that $\|v_k^\prime\|_\infty = O(1)$
imply that $\| \ket{V_k^{(N)}} \| = \sqrt{N}(1+O(h_N))$. Hence, the sum above is 
equal to 
\begin{eqnarray}
\frac{1}{N} \sum_{j=0}^{N-1} | v_k((j+1)h_{N}) (1 + O(h_N)) \nonumber \\
\quad\quad -  v_k((f(j)+1)h_{N_0}) (1 + O(h_{N_0})) |^2. \label{uk}
\end{eqnarray}  
Since $v_k(\cdot)$ is continuous with a bounded first derivative, we have that
\begin{equation}
v_k(x_{2,j}) = v_k(x_{1,j}) + O(|x_{2,j}-x_{1,j}|), \label{meanvalue}
\end{equation}
where $x_{1,j}=(j+1)h_N$ and $x_{2,j}=(f(j)+1)h_{N_0}$, 
$j=0,\ldots,N-1$. Clearly $|x_{2,j}-x_{1,j}| = O(h_{N_0})$.
Using (\ref{uk}), (\ref{meanvalue}) and the triangle inequality, 
we obtain from (\ref{sumestimate}) that
\begin{eqnarray}
&& \left\| \ket{U_k^{(N)}} - \ket{\tilde{U}_k^{(N)}} \right\| 
\leq O(h_{N_0})\frac{\| \ket{V_k^{(N)}} \|}{\sqrt{N}} + O(h_{N_0}) \nonumber \\
&& \quad  + O(h_{N_0}^q) = O(h_{N_0}^{\min \{1,q \}}). \label{failure}
\end{eqnarray} 
Hence, the probability of failure is bounded from above by 
$O(N_0^{-\min\{2,2q\}})$. It depends only on the order of convergence 
to the continuous problem and the number of points in the classically 
solved small problem. We can select an $N_0$ such that the probability 
of failure is less than 1/2, no matter how much larger $N$ is.
By choosing a large $N$, we can make the discretization error arbitrarily 
small. Equation (\ref{failure}) implies that the probability of obtaining 
the eigenvalue 
$e^{2\pi i \varphi_k}$ with accuracy $2^{-b}$ is at least 
$\frac{8}{\pi^2}(1-O(N_0^{-\min\{2,2q\}}))$.

We remark that any classical numerical algorithm that computes an 
eigenvalue, satisfying a specific (nontrivial) property, 
of a $N \times N$ 
unitary matrix takes time $\Omega(N)$. For example, one may want to find 
the eigenvalue that corresponds to the ground state.
This is true even if a matrix is sparse and regardless of whether the 
algorithm is deterministic or randomized.
It is merely a consequence of the fact that the algorithm needs to consider 
all the (nonzero) elements of the matrix and there are at least $\Omega(N)$
of them. Alternatively, in the restricted case when the matrix is diagonal
finding one of its elements is a problem at least as hard as
searching an unordered list. The lower bound for searching yields the lower 
bound in our case. 

In conclusion, our method provides a highly efficient preparation of 
initial states for eigenvalue approximation, requiring only 
a small number of Hadamard gates. 
Thus the algorithm of Abrams and Lloyd, using our initial state, computes 
the eigenvalue exponentially faster than any classical algorithm.
The method can be generalized to higher 
dimensional continuous problems. This will be the subject of a future
paper.

\begin{acknowledgments}
This research was supported in part by the National Science
Foundation (NSF) and by the Defense Advanced Research Agency
(DARPA) and Air Force Research Laboratory under agreement
F30602-01-2-0523.
P. J. also acknowledges support from the foundation 
Blanceflor Boncompagni-Ludovisi, n\'ee Bildt.
We are grateful to  Joseph Traub and Arthur Werschulz 
for helpful discussions and valuable comments. 
\end{acknowledgments}

\bibliography{initial_state_preparation}

\end{document}